# High performance nanophotonic circuits based on partially buried horizontal slot waveguides

**Chi Xiong[1,2], Wolfram H.P. Pernice[1,2], Mo Li[1] and Hong X. Tang[1,*]**

[1] *Department of Electrical Engineering, Yale University, New Haven, CT 06511, USA*
[2] *These authors contributed equally to this work.*

*[*]Corresponding author: hong.tang@yale.edu*

**Abstract:** We present a novel platform to construct high-performance nanophotonic devices in low refractive index dielectric films at telecoms wavelengths. The formation of horizontal slots by PECVD deposition of high index amorphous silicon provides a convenient and low-cost way to tailor nanophotonic devices to application needs. Low propagation loss of less than 2 dB/cm is obtained allowing us to fabricate optical resonators with measured high optical quality factors exceeding $10^5$. We design and experimentally demonstrate on-chip grating couplers to efficiently couple light into integrated circuitry with coupling loss of 4 dB and optical bandwidth exceeding 110 nm. The entire on-chip circuitry consisting of input/output couplers, Mach-Zehnder interferometers with high extinction ratio and ring, racetrack resonators are designed, fabricated and characterized.

**1. Introduction**

In recent years slot waveguides have been receiving increasing attention as a new class of waveguides for a multitude of applications [1-4]. In the context of optical waveguiding, a slot waveguide is composed of one or several narrow low refractive index regions enclosed between areas of high refractive index. The field discontinuity at interfaces between the high and low refractive index regions leads to strong field enhancement in the low index region, particularly close to the interfaces [5,6]. Initial slot waveguides were realized by fabricating nanophotonic waveguides with small separation, which creates vertical slots. Because of the field concentration near the waveguides sidewalls, these vertical slot waveguides often suffer from enhanced scattering loss and thus large propagation loss. The best reported loss for the quasi-TE mode in a vertical slot waveguide with a single slot of 50 nm or less is greater than $11.6 \pm 3.6$ dB/cm [7]. As a result, optical resonators involving vertical slots feature relatively low quality factors, which constrains their application in certain photonic applications.

Scattering losses can be significantly reduced by employing slot waveguides with a horizontal slot layer instead of a lithographically defined vertical slot. A horizontal slot structure including a horizontal low index slot layer can be fabricated by layered deposition or thermal oxidation [8,9]. The corresponding slot waveguide devices have virtually no fabrication constraints on slot thickness and can have very low scattering loss due to small surface or interface roughness for the fundamental slot mode. In addition, multiple slot configurations can be used in a horizontal slot waveguide concept to provide enhanced optical confinement in the low index slot region [6]. In this article we report on a complete nanophotonic architecture based on horizontal slot waveguides. Employing PECVD multi-layer deposition onto silicon-on-insulator substrates provides a flexible and cost-effective way to tailor the optical properties of the waveguiding structure for specific applications. We deposit an amorphous silicon top layer onto a lower-refractive index slot layer. Optical quality thin films of a-Si:H are deposited by plasma enhanced chemical vapor deposition (PECVD) at low temperatures, typically around 400°C [10-12]. At these temperatures a significant atomic percentage of hydrogen is incorporated in the material to fill dangling Si bonds [13] and reduce optical loss by absorption. The guiding slot layer can be chosen freely. Here we employ silicon nitride and silicon dioxide slot layers as exemplary materials. Differing from previous slot-designs, we pattern waveguiding geometry solely in the top silicon layer. By etching only the silicon top layer, slotted ridge waveguides are formed and scattering loss in the waveguides is minimized.

The horizontal slot waveguide provides a convenient way to realize efficient grating couplers. Grating couplers are emerging as a promising way to couple light into integrated photonic circuits. They allow for coupling light into and out of on-chip devices through first order Bragg reflection. As a result, optical chips can be readout from the top, which alleviates precision alignment significantly. It is generally the case that efficient grating couplers require a shallow grating structure to achieve high coupling efficiency. Our slot waveguide design lends itself naturally to shallow grating design, because only the top layer is etched and thus the grating is inscribed only into the top layer. In addition to grating couplers we realize on-chip photonic circuitry. We fabricate Mach-Zehnder interferometers with high extinction ratio and ring/racetrack resonators. We achieve propagation loss of less than 2 dB/cm and loaded quality factors in our optical cavities exceed $10^5$. These are the highest $Q$ factors ever realized in slot waveguides to date. Our platform holds promise for embedding functional materials that have low refractive index on a silicon photonics platform.

## 2. Fabrication and design of the slot waveguides

Our photonic platform is realized on commercial Silicon-On-Insulator (SOI) wafers, Soitec smart-cut with a buried oxide (BOX) layer of 3 μm thickness. The top silicon layer is thinned down to 110 nm thickness by thermal oxidization and subsequent wet-etch. The dielectric slot layer is deposited by PECVD in silane/nitrous oxide (for silicon dioxide slot) and silane/ammonia (for silicon nitride slot) chemistry with a targeted thickness of 80 nm. The final waveguiding layer composed of amorphous silicon (a-Si) is deposited by PECVD in a 150 sccm flow of silane precursor at 400°C at 2.4 torr. The vertical layer geometry used throughout the paper is shown schematically in Fig.1a). After the deposition process nanophotonic structures are defined by electron-beam lithography and subsequent reactive ion etching (RIE) using chlorine gas chemistry.

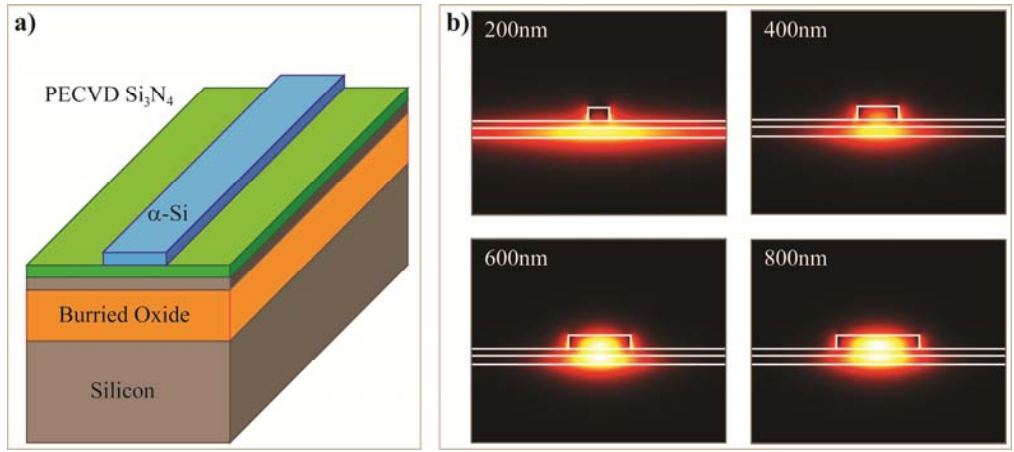

Fig.1 a) The vertical layer structure used to define the partially buried horizontal slot waveguides. PECVD silicon nitride (green) and amorphous silicon (blue) layers are deposited onto SOI substrates. b) Calculated mode profiles for horizontal slot waveguides with different waveguide widths ranging from 200 nm to 800 nm. Increasing the width of the top waveguide confines the mode laterally to the slot region.

We consider nanophotonic waveguides of varying width, ranging from 200 nm to 900 nm. As shown in the mode profiles in Fig.1b) narrow waveguides support a fundamental mode predominantly guided in the bottom silicon layer. Increasing the waveguide width leads to improved confinement of the total mode to the slot region below the a-Si waveguide.

## 3. On-chip grating couplers

In order to provide convenient access to the fabricated photonic circuits we employ focusing grating couplers. Due to first-order Bragg reflection on the grating structure, light can be coupled into and out-of the photonic circuits into the vertical direction [14]. Thus optical access to the chip is achieved by vertical alignment, which is much easier than conventional optical alignment through butt-coupling with lensed fibers or inverse tapers. Grating couplers have the advantage of not requiring polished facets for coupling, which enables wafer scale testing of the integrated circuits. In addition, grating couplers are very compact and have a large optical bandwidth [15-17].

Our photonic platform is designed for high-yield and easy fabrication, thus only a single electron-beam lithography and RIE step is desired. In traditional SOI nanophotonic circuits this approach will results in deep etched grating couplers. Such grating couplers do not allow for good coupling efficiency. The reported fiber coupling efficiency obtained with standard uniform grating structures amounts to roughly 20% [18]. In order to improve the coupling efficiency, one common approach is to employ shallow etch in the grating area. Shallow etching improves the coupling efficiency significantly. Our horizontal slot design lends itself naturally to a shallow grating etching, because only the a-Si layer is removed during the one-step etching.

We design grating couplers employing the finite-difference time-domain method. Because the grating structure is translational invariant the devices are investigated in a two-dimensional cross-section through the grating structure. The couplers are designed for operation at 1550 nm by adjusting the grating period. As shown in Fig.2a) the coupling loss can be as low as 3.4 dB from the simulation results. From the simulation results we find and optimized coupler period of 710 nm to obtain a central coupling wavelength of 1550 nm In particular, the coupler bandwidth measured at the 3 dB points is greater than 55 nm.

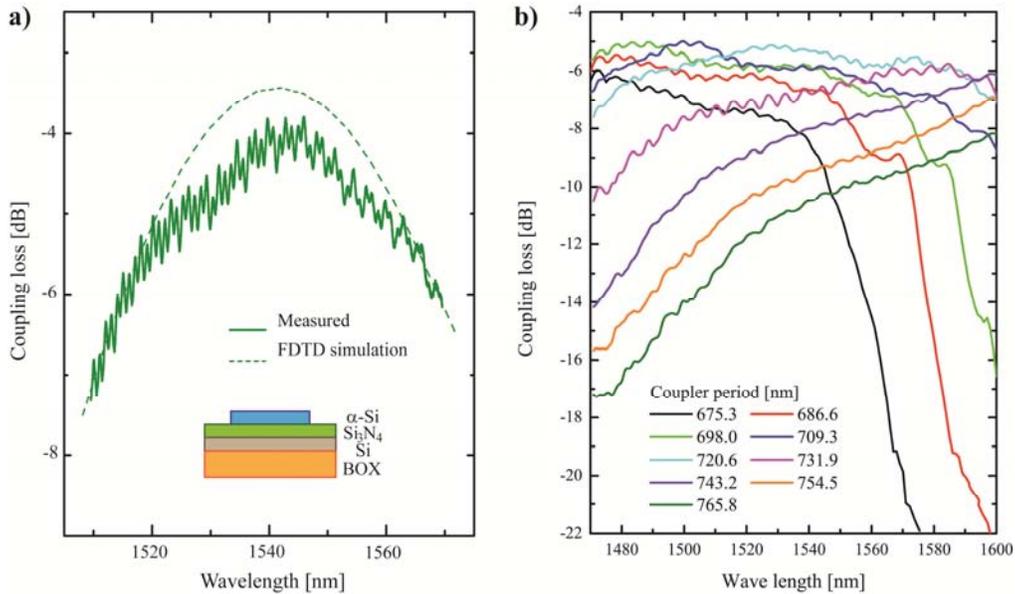

Fig.2 a) Simulated and measured transmission profile of a $SiN_x$ slot grating coupler with a central coupling wavelength of 1540 nm for a period of 710 nm. The nitride slot layer is 80 nm thick. The coupler features a minimum coupling loss of ~4 dB and a 3 dB bandwidth exceeding 55 nm. b) The measured transmission profiles of a $SiO_2$ slot grating coupler with a coupler period ranging from 675 nm to 754 nm. The oxide slot layer is 80nm thick. The coupler features a minimum coupling loss of ~5 and a 3 dB bandwidth exceeding 110 nm. Notable is the wide flat-top regime in the central coupling window.

In order to verify the theoretical predictions we fabricate grating couplers from slot substrates with a a-Si layer of 110 nm thickness and a silicon nitride slot layer of 80 nm thickness. As shown in Fig.2a) the measured transmission profile follows the simulated transmission curve well. The measured coupling loss is 4 dB, corresponding to 40% coupling efficiency. Discrepancy to the theoretical result is expected due to fabrication imperfections and the propagation loss of the waveguides within in the device.

In addition to the couplers built on a silicon nitride slot layer we also fabricate grating couplers with a silicon dioxide PECVD slot layer. Oxide based grating couplers feature a higher coupling bandwidth as shown in Fig.2b). We display the measured coupler profiles for different grating periods, shifting the central coupling wavelength to longer wavelengths with increasing coupler period. Striking is the wide coupling bandwidth, which spans beyond the tuning range of our laser. Compared to the nitride grating coupler the coupling loss is slightly higher (5 dB) at the central coupling point. From the measured result we estimate a 3 dB bandwidth of more than 110 nm for the couplers fitting into the tuning range of the laser. The increased optical bandwidth observed for the silicon oxide slot couplers is due to the reduced effective index of the coupling structure. This leads to improved mode matching properties of the coupling region over a wider wavelength range. In addition, the coupler features a flat passband over a wide wavelength range, which makes the design particularly useful for broadband spectroscopic applications. The coupling bandwidth is much improved compared to traditional SOI couplers, which feature a reduced 3 dB bandwidth of roughly 30 nm [19].

**4. Mach-Zehnder interferometers**

The grating couplers described above are employed to provide optical input and output ports to on-chip photonic circuitry. We first investigate Mach-Zehnder interferometers (MZIs), which are commonly used for sensitive metrological applications and nowadays also for the measurement of extended photonic functionalities [19].

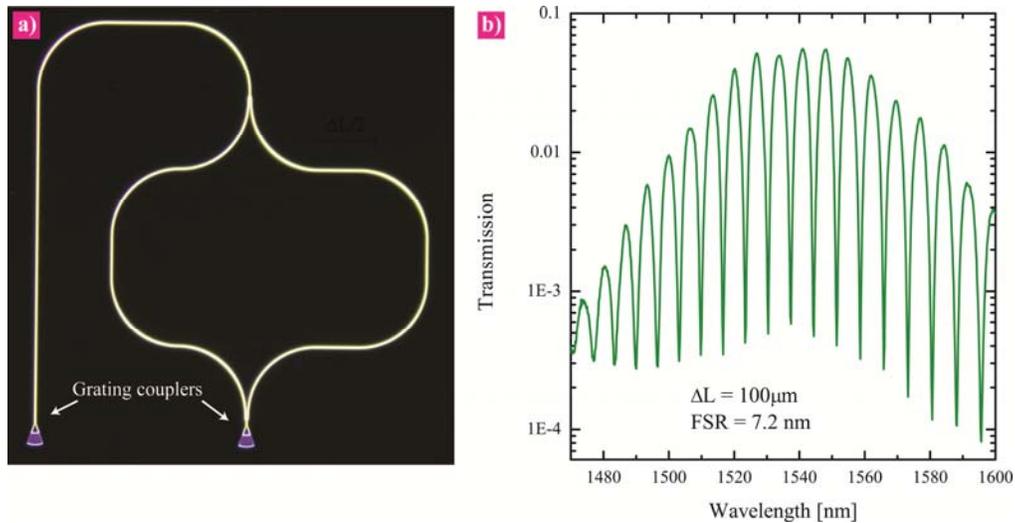

Fig.4 a) Optical image of a fabricated photonic circuit with input/output grating couplers and an integrated Mach-Zehnder interferometer with a path difference of 100 μm. The waveguide width is 900 nm. The horizontal slot is made of 80 nm PECVD $SiN_x$. b) The measured response of the fabricated sample shown in a). The response shows the typical interference fringes of an MZI, enveloped by the profile of the grating coupler. The free-spectral range is 7.2nm, which implies a waveguide group index of 3.35. The extinction ratio is greater 20 dB which illustrates that the arms of the MZI are well balanced.

In the MZIs an optical waveguide is split into two separate arms, which are joined together after a chosen propagation distance. A given path difference between the arms leads to a characteristic interference pattern at the output of the interferometer due to constructive and destructive interference. The free spectral range (FSR) of the interferometer is determined by the path difference and the group index of the waveguide. When the optical loss in both arms is equal and the splitting ratio is 50:50, the total loss through the interferometer is small and the extinction ratio between the peaks and the valleys of the fringes is high. In this case the interferometer is considered to be balanced.

We fabricate MZIs from silicon nitride horizontal slot waveguides with a path difference of 100 μm. The waveguide width is fixed at 900 nm, while the thickness of the nitride slot layer is kept at 80 nm. Given the group index of the waveguide of 3.35 obtained from finite-element simulations, the FSR of the MZI amounts to 7.2 nm. Therefore many fringes will fit into the bandwidth of the grating couplers. In Fig.4a) we show an optical micrograph of a fabricated sample. The image illustrates the layout of the photonic circuit, where the MZI is enclosed between the two focusing grating couplers. The measured response of the device is shown in Fig.4b), featuring nice interference fringes of a MZI. The response is enveloped by the profile of the grating couplers. From the measurement we obtain an extinction ratio of more than 20 dB, which illustrates that the interferometer is well balanced. The transmission loss at the peak of the interference fringes compared to a calibration sample without the interferometer amounts to 1 dB, which further proofs that the two arms of the MZI are balanced and the splitting ratio at the input and output is indeed close to 50:50.

## 5. Optical resonators

In addition to MZI devices we also fabricate resonant optical cavities on chip. Because we are investigating waveguide based geometries we focus here on ring and racetrack resonators. We fabricate samples with different coupling gaps in order to achieve critical coupling as shown in Fig.5. The samples are investigated optically through transmission measurements. First we analyze ring resonators with a radius of 100 μm and a waveguide width of 900 nm, fabricated with a silicon nitride slot layer. An optical micrograph of a fabricated device is shown in the inset of Fig.5a). Maximum extinction ratio in the transmitted optical signal is achieved under critical coupling conditions, which are reached when the optical power coupled into the ring matches the power dissipated inside the ring during one round trip. Under critical coupling conditions the measured loaded optical $Q$ corresponds to half the intrinsic optical quality factor. Under near-critical coupling condition with a coupling gap of 300 nm we find an extinction ratio of ~20 dB, as shown in the transmission spectrum in Fig.5a). From fitting the resonance dips with a Lorentzian curve we extract loaded optical $Q$ factors of 79,000, as shown in the zoom-in graph in Fig.5b). When measuring devices with larger coupling gap of 600 nm the devices are operated in the under-coupled regime and therefore the extinction ratio is smaller. For the best devices we find a cavity $Q$ of ~100,000. For comparison we also investigate ring resonators with a silicon dioxide slot layer. The transmission profile of a near critical coupled device is shown in Fig.5c). The flat top response reveals the broad bandwidth of the grating couplers. In this device good extinction of 20 dB is found, comparable to the silicon nitride slot device.

In order to establish the low propagation loss we consider a further set of ring resonators that are separated from the input waveguide by a larger gap of 500 nm. When measuring these weakly coupled devices, we find high optical $Q$ of 125,000 as shown in Fig.5d). This measured $Q$ is the highest $Q$-factor obtained in a slot resonator to the best of our knowledge.

The measurements of the resonators allow us to extract the properties of the resonator in more detail, using the method introduced in references [20,21]. We define the minimum power transmission in the through-port of the ring as $\gamma$ and the -3dB bandwidth as $\delta\lambda$. The response period of the resonator is the FSR introduced in the previous sections. The waveguide power coupling coefficient can then be calculated to as $\kappa^2 = \pi\delta\lambda \times (1-\sqrt{\gamma})/FSR$, and the propagation power loss coefficient $\kappa_p$ (including the bending loss and radiation loss due to sidewall roughness) is given by $\kappa_p^2 = 2\pi\delta\lambda\sqrt{\gamma}/FSR$ [21]. To be compared with the losses in straight waveguides quoted in dB/cm, the propagation loss in a microring resonator can be expressed as $\alpha_{dB} = -10 \times \log_{10}(1-\kappa_p^2)/2\pi R$, where $2\pi R$ is the perimeter of the microring resonator. The total quality factor of the ring is then defined as $Q_t = \lambda/\delta\lambda = 2\pi\lambda/(FSR \times (2\kappa^2 + \kappa_p^2))$. Using the above expressions the intrinsic quality factor is then given as $Q_i = 2\pi\lambda/FSR \times \kappa_p^2 = Q_t/\sqrt{\gamma}$.

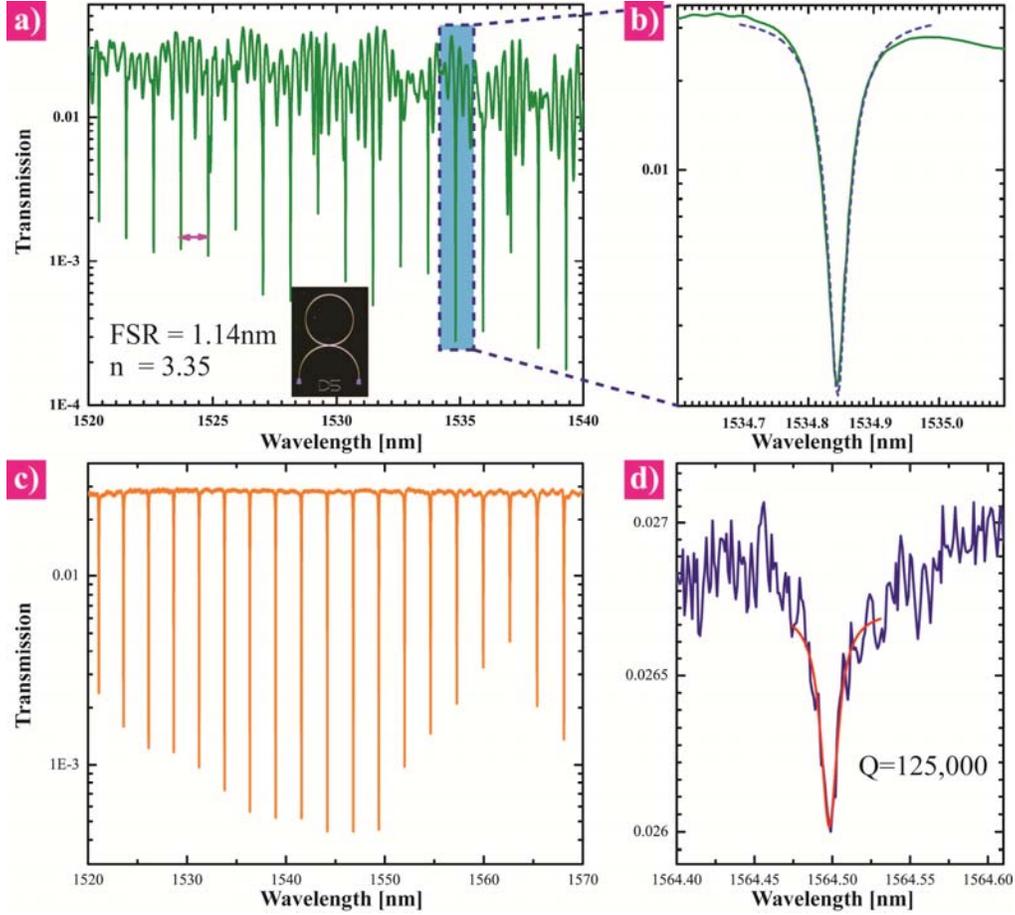

Fig.5 a) The transmission profile of a ring resonator fabricated from a silicon nitride horizontal slot waveguide. The fabricated device is shown in the optical micrograph in the inset. b) Zoom into one of the ring resonances for the near critical coupling case. A $Q$ of 79,000 is found from the fit with a Lorentzian dip. c) The transmission response of a near critical coupled ring resonator with a silicon dioxide slot layer, showing an extinction ratio of 20dB. The broadband flat top response is the envelope of the slot grating couplers. d) A high-$Q$ resonance measured in a separate, weakly coupled ring resonator with gap of 500nm. The Lorentzian fit reveals a best optical $Q$ of 125,000.

Given the extinction ratio for the device in Fig.5b) of 12.43 dB ($\gamma= 0.0571$), $\delta\lambda$=19.5 pm and FSR=1.12 nm, we find a total $Q_t$ of 79,000 at 1534.84 nm. The extracted power loss coefficient is then $\kappa_p$=0.1616, which implies a propagation loss of 1.83 dB/cm. The corresponding intrinsic $Q_i$ is $3.3\times10^5$ and the coupling coefficient is determined to be $\kappa$=0.204.

For a typical resonance in Fig.5c) for the oxide ring resonator we find an extinction ratio of 17.99 dB ($\gamma= 0.0159$), $\delta\lambda$=50.5 pm and FSR=2.63 nm, we find a total $Q_t$ of 30,000 at 1554.68 nm. The extracted power loss coefficient is then $\kappa_p$=0.1229, which implies a propagation loss of 2.1 dB/cm, given the ring radius of 50 μm. The corresponding intrinsic $Q_i$ is $2.44\times10^5$ and the coupling coefficient is determined to be $\kappa$=0.23.

In addition to ring resonators we investigate racetrack resonators with a straight coupling section of 100 μm length, as shown in Fig.6. Racetrack resonators can be considered extended ring resonators, into which two straight waveguide sections have been inserted. The resonators are coupled to an input waveguide, which is separated from the resonator by a coupling gap *g*. The straight coupling section presents a directional coupler. Therefore the coupling ratio into the waveguide depends on the optical wavelength. When the length of the straight section is long enough to be close to the coupling length of the directional coupler, the critical coupling condition for the racetrack resonator can always be matched for a given wavelength. This is observed in the transmission spectrum in Fig.6a). Shown is the measured transmission for silicon nitride slot device. A zoom into the resonances reveals an optical *Q* of 47,000, as shown in Fig.6b). Compared to the ring resonator the *Q* is slightly reduced due to the increased resonator length and thus enhanced round trip loss.

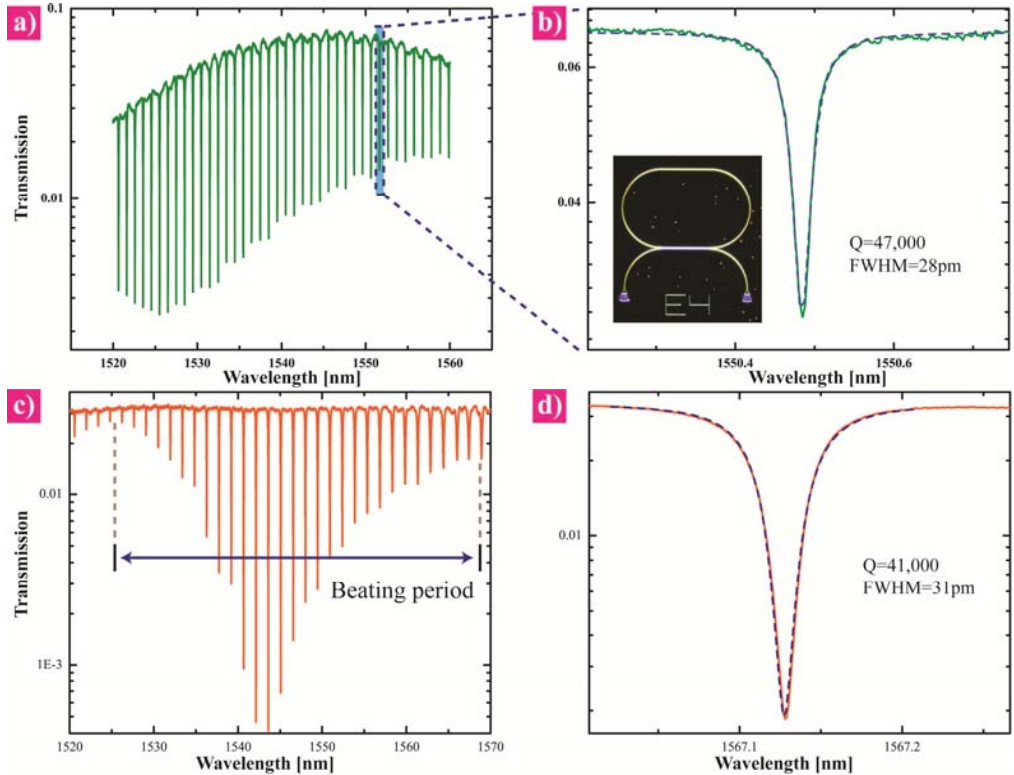

Fig.6 a) Shown is the measured transmission response of a racetrack resonator with a silicon nitride slot layer. The waveguide width is 900 nm. The transmission profile is enveloped by the coupler response, whereas the extinction ratio is enveloped by the beating pattern of the input directional coupler. b) The fitted response shows an optical Q of 47,000, slightly reduced from the Q measured in ring resonators. c) The transmission profile for a racetrack resonator fabricated with a silicon dioxide slot layer. The waveguide width is 900 nm. The beating pattern reveals a beating period of roughly 43 nm. d) The fitted response of one of the resonances, showing an optical *Q* of 41,000, comparable to the *Q* found in the nitride slot resonators.

Similar behavior is observed for racetrack resonators fabricated with a silicon dioxide slot layer, as shown in Fig.6c), d). In Fig.6c), the beating pattern due to the straight coupling section is clearly visible, with a beating period of roughly 43 nm. Taking into account the measured group index of the waveguide of 3.6, the beating length is close to the numerically expected beating period of 46 nm. By fitting the response in Fig.6d) to a Lorentzian we find a maximum *Q* of 41,000, which is comparable to the *Q* found for the nitride slot resonator. Under optimal coupling conditions the extinction ratio for resonances around 1545 nm is close to 20 dB, which also corresponds to the critical coupling extinction found in the ring resonators in the previous section.

## 6. Conclusion

In conclusion we have presented a flexible nano-photonic platform based on partially buried horizontal slot waveguides. Our design provides engineering flexibility in both slot design and photonic component engineering. The slot structure can be designed for almost all materials with the refractive index lower than that of silicon, which is the case for many functional dielectric materials. We have realized photonic circuits on this platform using two exemplary slot materials, i.e. silicon nitride and silicon dioxide. Because we only etch the top silicon layer, scattering loss from waveguiding side-walls is reduced and we achieve propagation loss as low as 2 dB/cm. The low propagation loss allows us to realize high loaded optical $Q$ factors exceeding $10^5$. From the measured $Q$ we extract intrinsic optical $Q$ factors as high as $3.3 \times 10^5$. Comparable propagation loss on the order of 2dB/cm is achieved for both material systems. By improving lithography and etching of the top silicon layer $Q$ factors up to a million will be feasible.

## Acknowledgements

This work was supported by a seedling program from DARPA/MTO and the DARPA/MTO ORCHID program through a grant from AFOSR. H.X.T acknowledges support from a Packard Fellowship in Science and Engineering and a CAREER award from the National Science Foundation. W.H.P. Pernice would like to thank the Alexander-von-Humboldt foundation for providing a postdoctoral fellowship. We are grateful to Michael Rooks for his help with the e-beam lithography. Electron beam lithography was carried out at the Center for Functional Nanomaterials, Brookhaven National Laboratory, which is supported by the U.S. Department of Energy, Office of Basic Energy Sciences, under Contract No. DE-AC02-98CH10886.